\documentclass{article}

\usepackage{microtype}
\usepackage{graphicx}
\usepackage{subfigure}
\usepackage{booktabs} 

\usepackage{hyperref}
\usepackage[table,xcdraw]{xcolor}

\usepackage[accepted]{icml2023}

\usepackage{amsmath}
\usepackage{amssymb}
\usepackage{mathtools}
\usepackage{amsthm}

\usepackage[capitalize,noabbrev]{cleveref}

\theoremstyle{plain}

\theoremstyle{definition}

\theoremstyle{remark}

\usepackage[textsize=tiny]{todonotes}

\icmltitlerunning{Fisher-Weighted Merge of Contrastive Learning Models in Sequential Recommendation}

\begin{document}

\twocolumn[
    \icmltitle{Fisher-Weighted Merge of Contrastive Learning Models \\ in Sequential Recommendation}



\icmlsetsymbol{equal}{*}

\begin{icmlauthorlist}
\icmlauthor{Jung Hyun Ryu}{equal,schai}
\icmlauthor{Jaeheyoung Jeon}{equal,schmath}
\icmlauthor{Jewoong Cho}{equal,schmath}
\icmlauthor{Myungjoo Kang}{schai,schmath}
\end{icmlauthorlist}

\icmlaffiliation{schai}{Interdisciplinary Program in Artificial Intelligence, Seoul National University, Seoul, Korea}
\icmlaffiliation{schmath}{Department of Mathematics, Seoul National University, Seoul, Korea}

\icmlcorrespondingauthor{Myungjoo Kang}{mkang@snu.ac.kr}

\icmlkeywords{Machine Learning, ICML}

\vskip 0.3in
]



\printAffiliationsAndNotice{\icmlEqualContribution} 

\begin{abstract}
Along with the exponential growth of online platforms and services, recommendation systems have become essential for identifying relevant items based on user preferences.
The domain of sequential recommendation aims to capture evolving user preferences over time.
To address dynamic preference, various contrastive learning methods have been proposed to target data sparsity, a challenge in recommendation systems due to the limited user-item interactions.
In this paper, we are the first to apply the Fisher-Merging method to Sequential Recommendation, addressing and resolving practical challenges associated with it.
This approach ensures robust fine-tuning by merging the parameters of multiple models, resulting in improved overall performance.
Through extensive experiments, we demonstrate the effectiveness of our proposed methods, highlighting their potential to advance the state-of-the-art in sequential learning and recommendation systems.
\end{abstract}

\begin{figure}[t]
\vskip 0.2in
\begin{center}
\centerline{\includegraphics[width=\columnwidth]{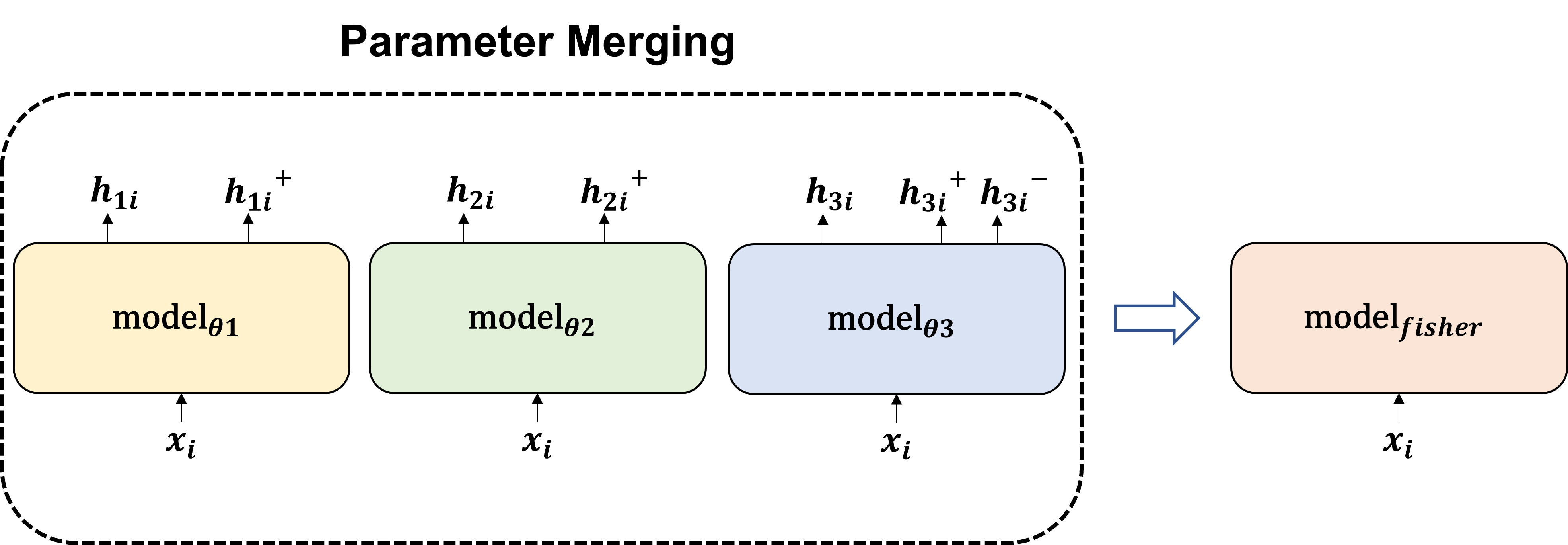}}
\caption{Overview of Parameter Merging.
Sharing model architecture, each models differ in how the contrastive loss is constructed. We merge models by using a weighted sum, where the weights are determined based on the posterior distribution of each model's parameters, assuming a Gaussian distribution.}
\label{ensemble_overview}
\end{center}
\vskip -0.2in
\end{figure}

\section{Introduction}
\label{introduction}
With the exponential growth of online platforms and services, a significant amount of data is being generated daily.
Recommendation systems have become crucial in utilizing this data effectively.
The system aim to identify relevant items based on user preferences and interests.
As user preferences evolve over time, sequential recommendation has gained attention as a subfield in this area.
We address the problem of sequential recommendation as follows.

Let $\mathcal{U}$ be the set of users $\mathcal{U} = \{u_1, u_2, \cdots, u_{|\mathcal{U} |}\}$, and $\mathcal{V}$ be the set of items as $\mathcal{V} = \{v_1, v_2, \cdots, v_{|\mathcal{V} |}\}$.
The sequence of user-item interaction for $u_i$ is a list with chronological order, $s_i=[v_1^{u_i}, v_2^{u_i}, \cdots, v_t^{u_i}, \cdots, v_{n_{u_i}}^{u_i}]$.
Here user $u_i \in \mathcal{U}$, $v_t^{u_i} \in \mathcal{V}$, and user $u_i$ interact item $v_t^{u_i}$ in time step $t$.
The length of sequence for user $u_i$ is $n_{u_i}$, and our object is to build a model predicting the item with which user is interact in the next time step, i.e. 
\begin{equation} p\left(v_{n_{u_i}+1}^{u_i} = v \,\middle\vert\, s_i\right). \end{equation}

The previous methodologies typically employ similar model structures but utilize various learning frameworks \cite{xie2022contrastive, qiu2022contrastive}.
Prior research has shown that ensemble methods yield significant benefits when multiple learning frameworks are employed \cite{gontijo2021no}.

We propose a practical and feasible method to ensemble the parameters of models trained with different contrastive learning techniques in a sequential recommendation.
The purpose of this study is to effectively aggregate the obtained parameters $\theta$ in various learning frameworks and hyperparameter settings, building on previous research and experiments.

Assuming the posterior distribution of parameters $\theta_m$ for each $m$-th model, Sec \ref{method:understanding_model_ensemble}, we achieved more effective ensemble results.
This approach allowed us to capture the uncertainty associated with each model's parameter estimates and leverage this information to enhance the ensemble process.
By considering the posterior distributions, we were able to account for the variability in parameter values across different models and obtain a more robust and comprehensive ensemble outcome.

\section{Related Works}

Researchers have explored various ensemble methods, including bootstrapping, bagging, and boosting, to improve model performance \cite{ganaie2022ensemble, breiman1996bagging, breiman2001random, natekin2013gradient, liu2014facial}. Ovadia et al. \cite{ovadia2019can} demonstrated the accuracy of ensembles even in the presence of distribution shift, while Mustafa et al. \cite{mustafa2020deep} proposed a method that combines fine-tuned subsets of pre-trained models to achieve high accuracy and resilience to distribution shift. Parameter merging is another technique to reduce model size and computational requirements \cite{houlsby2019parameter}. However, ensemble methods often require additional training, which can be computationally expensive and time-consuming. \\

\subsection{Diverse Learning Framework}

Wenzel et al. \yrcite{wenzel2020hyperparameter} and Zaidi et al. \yrcite{zaidi2021neural} investigated the role of random hyperparamters and architectures in ensemble.
Gontijo et al. \yrcite{gontijo2021no} demonstrated the ensemble effect across various training methodologies; initialization, hyperparameter, architecture, framework, and dataset levels.
Diverse training methodologies exhibit different generalization capabilities, ultimately lead to uncorrelated errors.
Models tend to specialize in subdomains within the data and highlights the crucial role of ensemble techniques in enhancing overall performance.

\subsection{Merging Methods}
\paragraph{Model Soup} Model Soup \cite{wortsman2022model} presents an effective approach for combining parameters without additional training.
It demonstrates research findings that improve the performance of trained models by constructing a "recipe" composed of diverse models and averaging their parameters.
The study introduces three methods for creating the recipe: the uniform soup, which averages the parameter values of all models; the greedy soup, which sequentially adds models based on their performance ranking; and the learned soup, which identifies the optimal model interpolation through training.
These approaches contribute to enhancing the overall performance of the model without the need for additional training.
\paragraph{Fisher Merging} Within the scope of related works, parameter merging is interpreted as a process that maximizes the joint likelihood of model parameters' posteriors \cite{matena2022merging}. 
Previous study \cite{wortsman2022model} consider averaging as a scenario where the posteriors of these models are assumed to follow an isotropic Gaussian distribution, and the joint likelihood is maximized accordingly.
To refine this approach, efforts have been made to approximate the posterior of the model using Laplace approximation \cite{ daxberger2021laplace}.
In this case, the distribution of each model is modeled by assuming the mean as the observed, which can be interpreted as trained parameter and the variance as the Gaussian distribution's Fisher matrix. By employing this formulation, the joint likelihood is calculated. 

\subsection{Sequential Recommendation System}\label{sec:seq_rec}
SASRec \cite{kang2018self} employ Transformer layers to dynamically assign weights to previous items. 
BERT4Rec \cite{sun2019bert4rec} demonstrate an improvement by incorporating user behavior information from both directions using a bidirectional Transformer. 
CL4SRec \cite{xie2022contrastive} employed three data augmentation techniques, namely item cropping, item masking, and item reordering, to create pairs for contrastive learning.
DuoRec \cite{qiu2022contrastive} integrated two types of contrastive loss. 
Firstly, it incorporated unsupervised augmentation using dropout-based model-level augmentation to generate positive pairs. Secondly, it incorporated supervised positive sampling, which involves creating pairs by considering sequences with the same target item as positive samples.
\section{Methodology}

We perform model ensemble based on different types of loss functions. BERT4Rec \cite{sun2019bert4rec}, CL4SRec \cite{xie2022contrastive}, and DuoRec \cite{qiu2022contrastive} share the basic structure of BERT4Rec \cite{sun2019bert4rec}. 
However, they differ in the sense of constructing positive pairs, a key component of their learning framework of constrastive learning.

Figure \ref{ensemble_overview} represents the overview of parameter merging process.
By sharing the structure of the model, which is parameterized with diverse learning frameworks, we can leverage ensemble methods to our advantage.
Furthermore, inspired by previous studies demonstrating the effectiveness of ensemble models trained using various learning methods, we apply parameter merging techniques, namely Parameter Averaging and Fisher-weighted Parameter Merging, described in Section \ref{par:fisher_merging}, to combine these models.

\subsection{Understanding Model Ensemble}
\label{method:understanding_model_ensemble}
We follow the work of Matena et al \yrcite{matena2022merging}. Consider a scenario where we have models with the same structure, $\text{model}_1, \text{model}_2, \cdots, \text{model}_M$, with corresponding parameters $\theta_1, \theta_2, \cdots, \theta_M$. 
Our objective is to find the parameter $\theta^*$ that maximizes the joint likelihood of the posteriors of these parameters. 

The posterior of $\theta_m$ can be represented as $p\left(\theta | \theta_m\right)$.
Since obtaining this posterior directly is generally challenging, it can employ approximation methods such as Laplace approximation to make assumptions and seek the parameter $\theta^*$ \cite{mackay1992practical, daxberger2021laplace}.
Let us interpret the process of finding $\theta^*$ as maximizing the joint likelihood, $\sum_m \log p\left(\theta \mid \theta_m \right)$.
Assuming that $p\left(\theta \mid \theta_m\right)$ follows a Gaussian distribution, we set the mean of this Gaussian distribution as the observed $\theta_m$ and examine the procedure for averaging parameters and Fisher merging separately, depending on the method used to assume the variance.

\paragraph{Averaging Parameters} \label{par:avg_param}
Assume that the posterior $p\left(\theta \mid \theta_m\right)$ follows a Gaussian distribution $\mathcal{N}\left(\hat{\theta}_m, I\right)$.
Here, $\hat{\theta}_m$ represents the parameters of the trained $m$-th model, and $I$ denotes the identity matrix. 
In this case, the desired solution $\theta^*$ can be obtained as the average of the parameters of the candidate models, as shown in eq.\ref{eq:average_likelihood_solution}: 
\begin{equation} \label{eq:average_likelihood_solution}
\theta^* = \underset{\theta}{\arg\max} \sum_m \log p\left(\theta \mid \theta_m, I\right) = \frac{1}{M}\sum_m\theta_m .
\end{equation}

\paragraph{Fisher Merging} \label{par:fisher_merging}
Let us consider the posterior $p\left(\theta \mid \theta_m\right)$ following a Gaussian distribution $\mathcal{N}\left(\hat{\theta}_m, H^{-1}\right)$. 
Here, $\hat{\theta}_m$ represents the parameters of the trained $m$-th model, and $H$ corresponds to the Hessian matrix of $\theta_m$ obtained through the second-order Taylor expansion at the mode of the posterior. 
It has been established that the Hessian matrix in this distribution coincides with the Fisher information, but for computational efficiency, we only utilize the diagonal elements of the Fisher matrix \cite{kirkpatrick2017overcoming}.

The desired solution $\theta^*$ can be expressed as eq.\ref{eq:fihser_likelihood}, capturing the essence of the Fisher likelihood :
\begin{equation}\label{eq:fihser_likelihood}
\theta^* = \underset{\theta}{\arg\max} \sum_m \lambda_m \log p\left(\theta \mid \theta_m, F_m\right),
\end{equation}
where $F_m = \mathbb{E}_x\mathbb{E}_{y\sim p_\theta\left(y|x\right)}\nabla_\theta \log {p_\theta\left(y|x\right)} \nabla_\theta \log {p_\theta\left(y|x\right)}^T$. The closed-form solution for $\theta^*$ can be obtained as shown in eq.\ref{eq:fisher_solution}, which directly incorporates the Fisher matrix. In practice, we utilize an empirical estimate of the Fisher matrix, denoted as $\hat{F}$, as shown eq.\ref{eq:fisher_solution} \cite{kirkpatrick2017overcoming}.
\begin{equation} \label{eq:fisher_solution}
\theta^{*\left(j\right)} = \frac{\sum_m \lambda_m F_m^{\left(j\right)} \theta_m^{\left(j\right)}}{\sum_m \lambda_m F_m^{\left(j\right)}}, 
\end{equation} 
where $F_m = \frac{1}{N} \mathbb{E}_{y\sim p_\theta\left(y\mid x\right)} \left(\nabla_\theta \log {p_\theta\left(y\mid x\right)}\right)^2$ and $j= 1,\cdots,|\theta|$, considering as element-wise multiplication.

\subsection{Applying Model Ensemble}
By expressing the Fisher matrix we intend to compute in  eq.\ref{eq:fisher_solution} in terms of recommendation factors, we can decompose it into the following components:
\begin{equation}
\begin{alignedat}{3}
    &\mathbb{E}_{x_i}\mathbb{E}_{y\sim p_\theta\left(y\mid x_i\right)} \left(\nabla_\theta \log {p_\theta\left(y\mid x_i\right)}\right)^2 \\
    &=\frac{1}{N}\sum_i \sum_j p_\theta\left(y_j\mid x_i\right) \left(\nabla_\theta \log {p_\theta\left(y_j\mid x_i\right)}\right)^2 \\
    &=\frac{1}{|\mathcal{U}|}\sum_i^{|\mathcal{U}|} \sum_j^{|\mathcal{V}|}
    p_\theta\left(v_j\mid s_i\right) \left(\nabla_\theta \log {p_\theta\left(v_j\mid s_i\right)}\right)^2 .
\end{alignedat}
\end{equation}\label{eq:calculate_fisher}
There are two computational challenges associated with the above equation.
First, calculations need to be performed for each individual sample $s_i$.
Second, calculations need to be performed for each item $v_j$ within a single sample.
The reason why these points acts as a drawback in recommendation systems is due to the large number of users and items in the data.
For instance, in the case of MovieLens-1M dataset \cite{harper2015movielens}, there are about 6000 users and 3500 items.
However, performing Fisher matrix calculations that require differentiation with respect to $\theta$ for each user and item becomes a computational burden.

\subsubsection{Sampling sequences}
\paragraph{Batch-wise Computation}
To address the first challenge of performing computations on individual samples, we reinterpret the equation and carry out the calculations on a batch basis. It should be noted that $p_\theta\left(v_j|s_i\right)$ can vary for each sample $s_i$. Therefore, we perform the sorting of $p_\theta\left(v_j|s\right)$ to address this variation, where $\text{BS}$ indicates batch size: 
\begin{equation}\label{eq:batch_computation}
\begin{alignedat}{2}
&\sum_i^{|\mathcal{U}|} \sum_j^{|\mathcal{V}|} p_\theta\left(v_j\mid s_i\right) \left(\nabla_\theta \log {p_\theta\left(v_j\mid s_i\right)}\right)^2 \\
&=\sum_{\text{BS}_k} \sum_j^{|\mathcal{V}|} \left(\sum_i^{\text{BS}_k}p_\theta\left(v_j| s_i\right) \right) \left(\nabla_\theta \sum_i^{\text{BS}_k}\log {p_\theta\left(v_j|s_i\right)}\right)^2.
\end{alignedat}
\end{equation}

\subsubsection{Sampling items}
To alleviate the computational burden associated with iterating over all $j$ values, which scales with $|\mathcal{V}|$, we employ a sampling-based approach within the methodology.
This sampling strategy aims to reduce the computational cost while maintaining the representativeness of the calculations.

\paragraph{Random Sampling} \label{par:random_sample}
We compute the eq.\ref{eq:calculate_fisher} by randomly sampling $j$ from the total number of items. 
This process was performed to calculate the Fisher matrix without any specific assumptions or prior knowledge. 

\paragraph{Top-$k$ Sampling} \label{par:topk_sample}
The probability which is output by the model can be interpreted as the preference or likelihood of the recommended items for a given sample.
Based on this interpretation, we select a set of $n$ items that are most likely to be of interest to the corresponding user, i.e. $p_{\theta}\left(v_j\mid s_i\right)$. 
Subsequently, we compute the Fisher matrix with these selected items as the focal points.
By focusing on this subset of items that are expected to be of highest interest, we aim to capture the relevant information for optimizing the model's performance effectively.
\begin{equation}
\begin{alignedat}{2}
&\sum_j^{|\mathcal{V}|} p_\theta\left(v_j\mid s\right) \left(\nabla_\theta \log {p_\theta\left(v_j\mid s\right)}\right)^2\\
&\approx \sum_j^{\text{top-}k}p_\theta\left(v_j\mid s\right) \left(\nabla_\theta \log {p_\theta\left(v_j\mid s\right)}\right)^2.
\end{alignedat}
\end{equation}

\paragraph{Model-based Sampling} \label{par:yrate_sample}
To select a subset of items for further analysis, we randomly sampled items based on their conditional probability $p_\theta(v_j|s_i)$ using a weighted random selection process. The selection probability of each item was determined by its associated probability stored in the model's output. By selecting items with higher probabilities, we focused on a specific number of items that were more likely to align with the user's preferences or interests. This allowed us to analyze and evaluate the subset of items based on their associated probabilities obtained from the model's output. With $N$ denotes the sample size, his approximation can be represented as:
\begin{equation}
\begin{alignedat}{2}
&\mathbb{E}_{y\sim p_\theta\left(y\mid x\right)} \left(\nabla_\theta \log {p_\theta\left(y\mid x\right)}\right)^2 \\
&\approx \frac{1}{N}\sum_{v_j \sim p_\theta\left(v_j\mid s\right)}^{N}\left(\nabla_\theta \log {p_\theta\left(v_j\mid s\right)}\right)^2.
\end{alignedat}
\end{equation}

\paragraph{Calculate with target item} \label{par:target_sample}
We compute the Fisher matrix based on the target item, disregarding other items with limited direct relevance.
By employing this approach, we focus solely on the target item and its associated information to calculate the Fisher matrix.
Our rationale behind this decision is to prioritize the target item's impact on the model's optimization process, as it is directly linked to the specific objective or task at hand.
Consequently, we exclude items with minimal direct relevance to ensure a more targeted and meaningful computation of the Fisher matrix.
\begin{equation}
p_\theta\left(v_j^*\mid s\right) \left(\nabla_\theta \log {p_\theta \left(v_j^*\mid s\right)}\right)^2,
\end{equation}
where $v_j^*$ is the target item.

\section{Experiments}
We use MovieLens-1M dataset \cite{harper2015movielens} for experiments.
For each user, we have sequential data consisting of movies purchased in chronological order.
We adopt next-item prediction task (i.e. leave-one-out evaluation), following previous works \cite{sun2019bert4rec, xie2022contrastive, qiu2022contrastive}.
The last movie is considered as the test set, and the validation data is used to predict the preceding movies.
During training, we adopt a masked language modeling approach similar to BERT \cite{devlin2018bert}, where we mask certain movies in the sequentially ordered list and task the model with predicting them.

The evaluation method used in this study is the Normalized Discounted Cumulative Gain at 10 (NDCG@10), which is a ranking-based evaluation approach \cite{he2017neural}. It ranks the top 10 items predicted by the model based on their perceived preference and considers the actual ranking of the preferred items. A higher NDCG value, closer to 1, indicates better performance. Different NDCG values can be obtained depending on the selection of items, such as from the full item pool, a random set of 100 items, or the top 100 most popular items.

\subsection{Results of Model Merging}
Examine the results through Table \ref{tb:baseline_merge} and Table \ref{tb:fine-tune_merge}.
Table \ref{tb:baseline_merge} presents the results obtained by training models, namely BET4Rec \cite{sun2019bert4rec}, CL4SRec \cite{xie2022contrastive}, and DuoRec \cite{qiu2022contrastive}. We merge these models using Fisher methods.
While Table \ref{tb:baseline_merge} demonstrates the results of models trained solely from scratch.
Table \ref{tb:fine-tune_merge} represents the results of fine-tune setting.
We train the baseline model without contrastive loss for 20 epochs, which is the convergence point of the baseline experiment without any additional contrastive loss, similar to BERT4Rec.
Following this, each model; BERT4Rec \cite{sun2019bert4rec}, CL4SRec \cite{xie2022contrastive}, DuoRec \cite{qiu2022contrastive}, underwent fine-tuning according to their respective methods, and the results were merged using Fisher methods.
In both conditions, we fine-tuned addtional epoch after merging process.

Fisher merge fails to improve the performance of individual models in baseline setting.
When Fisher merge is applied during the fine-tuning setting, it leads to improved performance compared to individual models.
This finding aligns with previously reported phenomena \cite{ganaie2022ensemble} where individual models tend to achieve higher performance than merged model in the baseline setting.
However, the results of Fisher merge in the fine-tuning setting show comparable performance with the individual models in baseline setting, while the individual recipe models of fine-tuning setting do not exceed.
Also, the results indicate that even for models that have not been sufficiently trained such as CL4SRec in our setting, merging parameters resulted in comparable performance to other models, demonstrating robustness.

\begin{table}[h]
\caption{Results of parameter merging; Fisher-merge on Baseline Settings. The recipes used for merging were trained for the same number of epochs.
`POS.' refers to the method of constructing positive pair, `sup' to supervised augmentation and `-' to supervised and unsupervised augmentation in subsection \ref{sec:seq_rec}}
\label{tb:baseline_merge}
\vskip 0.15in
\begin{center}
\begin{small}
\begin{sc}
\begin{tabular}{cc|ccc}
\toprule
model    & pos   & full            & random          & popular         \\ 
\midrule
baseline &       & \textbf{0.1398} & \textbf{0.5651} & \textbf{0.0482} \\
cl4srec  &       & 0.0955          & 0.043           & 0.0429          \\
duorec   & sup   & 0.1348          & 0.5575          & 0.044           \\
         & unsup & 0.1301          & \underline{0.5592}    & 0.0438          \\
         & -     & \underline{0.1382}    & 0.5588          & \underline{0.0464}    \\ 
\midrule
\rowcolor[HTML]{EFEFEF} 
fisher   &       & 0.1289          & 0.5495          & 0.0472          \\ 
\bottomrule
\end{tabular}
\end{sc}
\end{small}
\end{center}
\vskip -0.1in
\end{table}

\begin{table}[h]
\caption{Results of parameter merging; Fisher-merge on fine-tune Settings. The recipes used for merging were trained on baseline model (without contrastive loss) and fine-tuned on each model.}
\label{tb:fine-tune_merge}
\vskip 0.15in
\begin{center}
\begin{small}
\begin{sc}
\begin{tabular}{cc|ccc}
\toprule
model    & pos   & full            & random          & popular         \\ 
\midrule
baseline &       & 0.135           & 0.5573          & 0.0426          \\
cl4srec  &       & 0.0585          & 0.0513          & \textbf{0.0466} \\
duorec   & sup   & 0.1346          & 0.5547          & \underline{0.0454}    \\
         & unsup & \underline{0.1358}    & \underline{0.5594}    & 0.0445          \\
         & -     & 0.1351          & 0.554           & 0.0423          \\ 
\midrule
\rowcolor[HTML]{EFEFEF} 
fisher   &       & \textbf{0.1386} & \textbf{0.5618} & 0.0428   \\
\bottomrule
\end{tabular}
\end{sc}
\end{small}
\end{center}
\vskip -0.1in
\end{table}



\begin{table*}[!ht]
\caption{Effect of Sampling Method and Sampling Size. We merge models in settings of Table \ref{tb:fine-tune_merge}, the fine-tune setting. We merged models with 4 sampling methods; random sampling, top-k sampling, model-based sampling, and calculate on target item, on 3 different sampling size; n=10, n=30 and n=50. \textbf{Bold} represents the best variant in each evaluation setting, and \underline{underlines} indicates the second best variation.}
\label{tb:sampling_method_table}
\vskip 0.15in
\begin{center}
\begin{small}
\begin{sc}
\begin{tabular}{cc|cc|cc|cc}
\toprule
                                      & sample size & \multicolumn{2}{c|}{FULL}         & \multicolumn{2}{c|}{RANDOM}       & \multicolumn{2}{c}{POPULAR}       \\ 
\midrule
                                      &             & NDCG10         & NDCG20         & NDCG10         & NDCG20         & NDCG10         & NDCG20         \\ 
\midrule
baseline                               &             & 0.135           & 0.1601          & 0.5573          & 0.5786          & 0.0426          & 0.0706          \\
CL4SRec                               &             & 0.0585          & 0.0751          & 0.0513          & 0.043           & \textbf{0.0466} & 0.0701          \\
DuoRec (sup.)                         &             & 0.1346          & 0.1591          & 0.5547          & 0.58            & 0.0454          & 0.068           \\
DuoRec (unsup.)                       &             & 0.1358          & 0.1609          & 0.5594          & 0.5782          & 0.0445          & 0.0742          \\
DuoRec (sup.\&unsup.)                 &             & 0.1351          & 0.1599          & 0.554           & 0.5732          & 0.0423          & 0.0724          \\ 
\midrule
random sampling      & 10          & 0.1379          & \underline{0.1638}    & 0.5606          & \underline{0.5825}    & 0.0457          & 0.0691          \\
                                      & 30          & 0.1366          & 0.1624          & 0.5584          & 0.58            & 0.0477          & \textbf{0.0726} \\
                                      & 50          & \underline{0.1386}    & 0.1636          & 0.5598          & 0.5813          & 0.0419          & 0.0419          \\ 
\midrule
top-k sampling       & 10          & 0.1364          & 0.1624          & 0.5602          & 0.5817          & 0.0446          & 0.0689          \\
                                      & 30          & 0.1373          & 0.1616          & \textbf{0.5637} & \textbf{0.5835} & 0.0457          & 0.0708          \\
                                      & 50          & \textbf{0.1387} & 0.1635          & 0.5592          & 0.5807          & 0.0424          & 0.0672          \\ 
\midrule
model-based sampling & 10          & 0.1358          & 0.1619          & 0.5564          & 0.5782          & 0.044           & 0.0696          \\
                                      & 30          & 0.1385          & \textbf{0.1646} & 0.5579          & 0.5784          & 0.0446          & 0.0689          \\
                                      & 50          & 0.138           & 0.1632          & 0.5605          & 0.5814          & \underline{0.0465}    & 0.0719          \\ 
\midrule
calculate on target item              & 1           & \underline{0.1386}    & 0.1628          & \underline{0.5618}    & 0.5806          & 0.0428          & \underline{0.0725}    \\ 
\bottomrule
\end{tabular}
\end{sc}
\end{small}
\end{center}
\vskip -0.1in
\end{table*}

\subsection{The Validity of Batch-wise Computation}
We performed batch-wise computations with the aim of implementing an efficient Fisher matrix calculation. Compared to computing on individual samples, grouping samples into batches allowed us to achieve computational efficiency.

The following Figure \ref{fig:50item_prob} in Appendix \ref{appendix:sorted_prob} illustrates the method for minimizing errors when performing calculations on a batch basis. The figure demonstrates that within a batch containing 10 samples, denoted as $s_i$, there is a phenomenon where the probabilities of item $v_j$ decrease in a similar manner. By sorting the samples $s_i$ based on the probability of $v_j$, even when grouping them into batches, it is possible to minimize the error described by the eq.\ref{eq:batch_computation}. Furthermore, the figure illustrates the rationale behind top-k sampling. For the top-k items, the probabilities hold meaningful information, whereas for the remaining items, the probabilities are nearly zero or close to it.

\subsection{Effect of Sampling Methods and Size}
To investigate the effect of sampling methods, we conduct experiments by varying the number of sampled samples and the sampling techniques employed.
Specifically, we consider three sample sizes: $n=10, n=30 \text{ and } n=50$, and four different sampling methods: random sampling, top-k sampling, model-based sampling, and calculate with target item.
The results of these experiments can be observed in Table \ref{tb:sampling_method_table}.
The table provides insights into the performance of each sampling method under different sample sizes, allowing us to analyze their respective effects on the task at hand.
Note that this result is calculated on batched data.
To examine the results of parameter merging, we conducted experiments in fine-tuning setting, explained in \ref{tb:fine-tune_merge}. 

The experiments revealed effective ensemble results, particularly showcasing the robust performance of CL4SRec \cite{xie2022contrastive}.
Despite having significantly lower performance compared to other models during the parameter merging process, the model with poor performance exhibited robust performance in the Fisher merge results.
Regarding the sampling methods, top-k sampling demonstrated the best performance.
This can be attributed to the concentration of probabilities assigned to specific items by the model, effectively approximating the Fisher criterion sought in the evaluation.
Also, the model-based sampling method exhibits a more pronounced improvement in performance as the sampling size increases compared to other models.
We interpret these results as being rooted in the direct interpretation of the equation defined for Fisher merging.
Interestingly, despite the fact that calculating Fisher matrix on target item has a single sample, the method demonstrated sampling effectiveness by achieving good performance even with a small sample size.
These findings shed light on the interpretation of experimental results in the context of deep learning research.

\begin{figure}[ht]
\begin{center}
\centerline{\includegraphics[width=0.8\columnwidth]{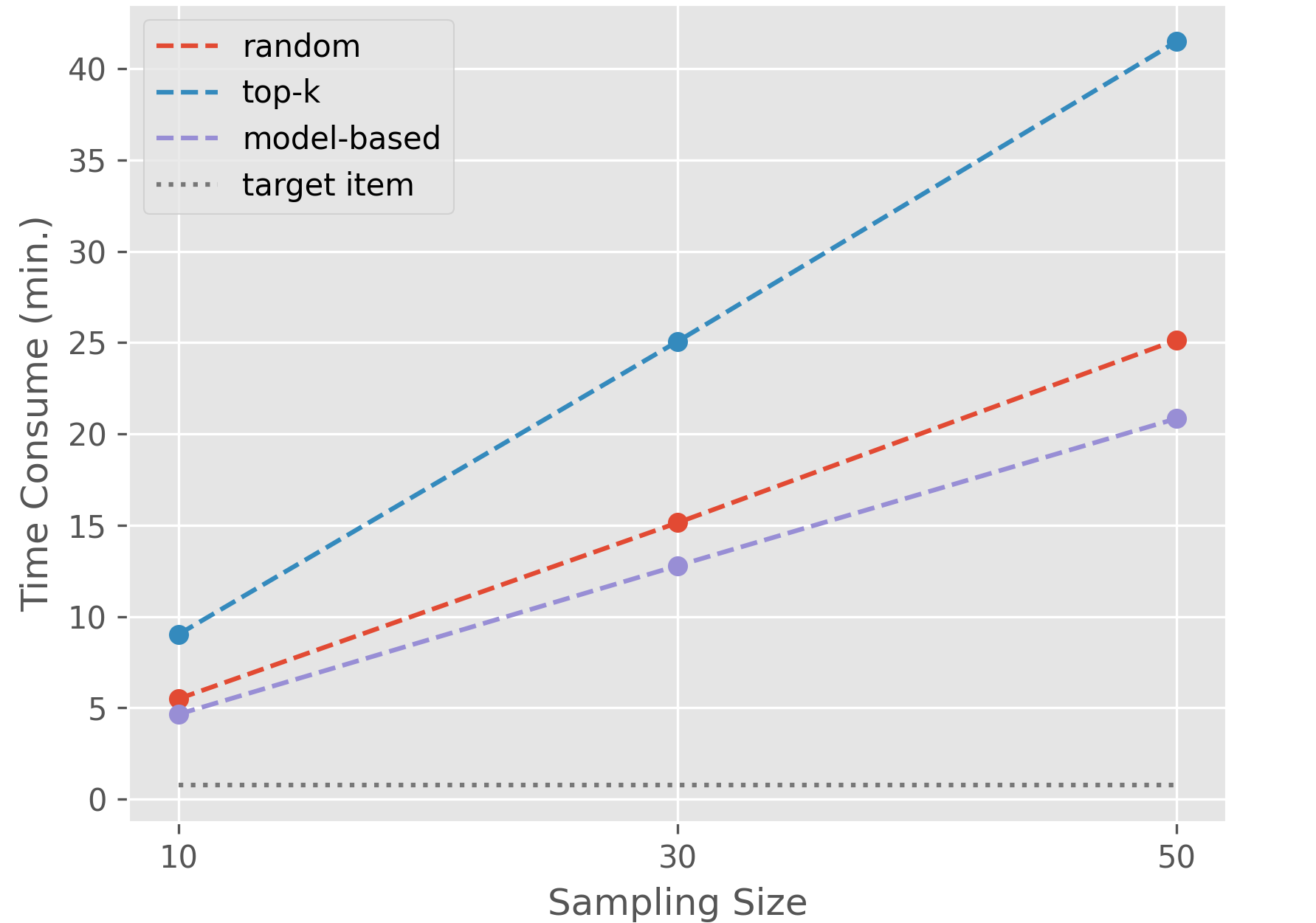}}
\caption{Measured Time Consumed for Each Sampling Method and Size. We sample items from MovieLens-1M dataset. Time is measured in batch-wise setting, where batch size is 256.}
\label{fig:comp_cost}
\end{center}
\end{figure}

\subsection{Computational Cost}
Figure \ref{fig:comp_cost} demonstrates computational cost in terms of time consumed during calculating Fisher matrix for single model.
The concept of parameter merging involves additional computation on the existing parameters.
Therefore, it is important to ensure efficiency in this process. To achieve efficiency, considerations such as calculating the Fisher matrix in batch units and performing sampling are necessary.
It is observed that, except for the calculation on the target item, the computational complexity increases linearly with the sampling size.
As for the calculation on the target item, the sampling size remains fixed at 1 since each sequence has a single target item.
Thus, our research is significant as it approximates the Fisher matrix calculation with a much smaller number of items (around 3000) compared to calculating it on the entire item set.

\begin{figure}[ht]
\begin{center}
\centerline{\includegraphics[width=\columnwidth]{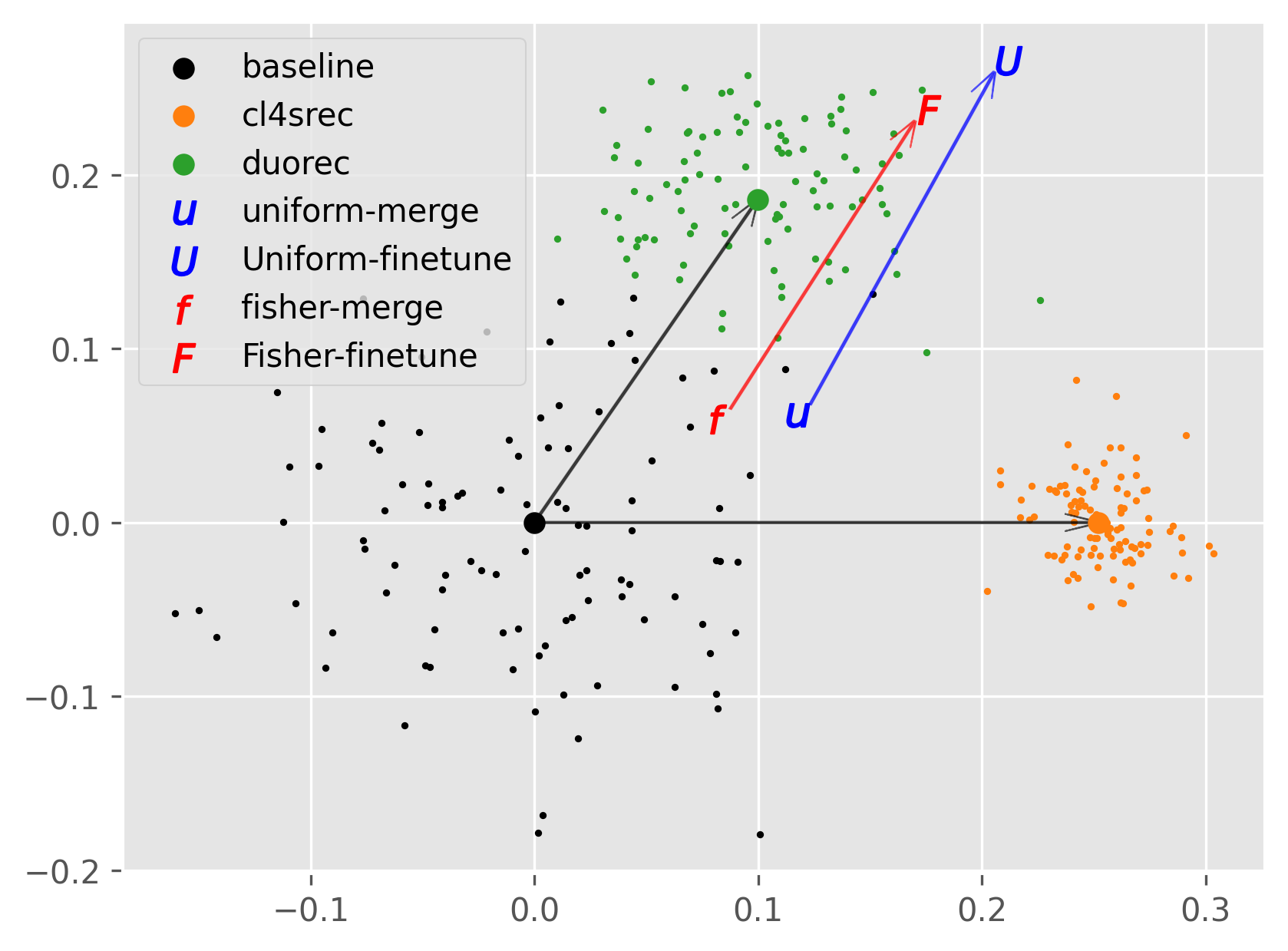}}
\caption{Visualization for Weights of Merged Models. Based on the plane containing 64-dimension parameters of three model, we visualised its weight, 100 samples from each posteriors and merged parameters}
\label{fig:visualize_weight}
\end{center}
\end{figure}

\subsection{Visualization of Merged Weights}
We present a visual illustration to aid in the intuitive understanding of the merged weights.
Figure \ref{fig:visualize_weight} represents the fine-tuning setting of \ref{tb:fine-tune_merge}, where the three centroids correspond to the weights of individual models.
The plane visualized in \ref{fig:visualize_weight} encompasses these three weights.
The scattered points, projected onto the plane, depict 100 samples drawn from $\mathcal{N}(\theta_m, F_m)$.
It is observed that the baseline weight exhibits the largest variance.
This can be attributed to the experimental setup where the baseline is pre-trained and then fine-tuned with CL4SRec \cite{xie2022contrastive} and DuoRec \cite{qiu2022contrastive}.
The weights obtained through uniform merging are represented as the average of the three centroid points, while the weights obtained through Fisher merging take into account the variances of these recipe weights.
It can be seen that the weights obtained through Fisher merging considered posterior and variance with Laplace approximation and provides nice initial point for fine-tune.


\section{Conclusion}
We apply ensemble technique, Fisher merging, for sequential models, enabling robust fine-tuning through parameter merging.
Our experimental results demonstrate the effectiveness of these proposed methods in improving recommendation performance.
These contributions have the potential to advance the field of sequential learning and recommendation systems, offering valuable insights for future research and practical applications.

\section*{Acknowledgements}
This work was support by the NRF grant [2012R1A2C3010887] and the MSIT/IITP [1711117093, 2021-0-00077, 2021-0-01343, Artificial Intelligence Graduate School Program(SNU)].

\nocite{langley00}

\bibliography{example_paper}
\bibliographystyle{icml2023}

\newpage
\appendix
\onecolumn
\section{Appendix}

\subsection{Motivation : Error Inconsistency}

\begin{table}[h]
\caption{Effect of Constructing Pair of Contrastive Loss (\%). We observe that models with divergent training methodologies exhibit distinct generalization behavior, resulting in highly uncorrelated errors.}
\label{tb:error_inconsistency}
\label{sample-table}
\vskip 0.15in
\begin{center}
\begin{small}
\begin{sc}
\begin{tabular}{c|ccc}
\toprule
            & similar &  & disssimilar \\
\midrule
CL4SRec \cite{xie2022contrastive}     & 8.05 & $<$   & 11.41     \\
DuoRec \cite{qiu2022contrastive}      & 8.67 & $<$   & 11.18     \\
\bottomrule
\end{tabular}
\end{sc}
\end{small}
\end{center}
\vskip -0.1in
\end{table}
Previous research \cite{gontijo2021no, yosinski2015understanding} demonstrated the increased effectiveness of ensemble methods as error inconsistency grows.
Building upon the existing research discourse, we conducted the current experiment.
In this study, we analyze the impact of Fisher merging in the context of sequential recommendation systems, attributing its effectiveness to the selection of recipe models trained using different frameworks .

In our experiments, we employ a model based on the BERT4Rec \cite{sun2019bert4rec} architecture as our baseline.
To enhance the performance of the model, we apply various data augmentation techniques to enable contrastive learning.

To analyze the effects of contrastive loss, we divide the training frameworks into two categories: similar and dissimilar.
The similar learning frameworks are trained using the same loss function but with slight variations such as different seeds and hyperparameters, indicating the relationship among models trained with small changes.
On the other hand, the dissimilar learning frameworks involve different data augmentation techniques, resulting in variations in the construction of positive and negative pairs for contrastive loss \cite{wang2020understanding}.

Error inconsistency \cite{geirhos2020beyond} refers to the percentage of data where two models have different classification results, with one model making correct predictions while the other model makes incorrect predictions.
Since we are not dealing with classification, we considered a model to have made a correct prediction if the value of NDCG@10 is above 0.5. 

By comparing the error inconsistency between similar framework and dissimilar framework, we observe the effectiveness of contrastive loss.
An observation that can be inferred from Table \ref{tb:error_inconsistency} is that the constructing positive pair for contrastive loss significantly affects the similarity of the samples that the models predict accurately.
As the method for constructing positive pair varies, the models demonstrate a considerable difference in their ability to predict samples correctly.
This finding highlights the sensitivity of the models to the specific construction of the contrastive loss, which in turn impacts their predictive performance.

\subsection{Robustness of Fisher Merging; Recipe Selection}
We compared two different recipe selection in Table \ref{tb:fine-tune_merge_abl}; Fisher merged parameters with least performance model and Fisher merged parameters without the model. In our experimental setup, CL4SRec did not exhibit superior performance compared to other models, considering the chosen hyperparameter settings and other factors. Therefore, we aim to leverage the elements of the recipe to demonstrate the robustness effect of Fisher merge. Our findings confirm that by removing underperforming models as individual components and applying Fisher merge, the resulting ensemble demonstrates robustness.
\begin{table}[h]
\caption{Ablation Study of Results of parameter merging; Fisher-merge on fine-tune Settings. The recipes used for merging were trained on baseline model (without contrastive loss) and fine-tuned on each model. Ablation Study illustrates the situation of recipe without the model with least performance.}
\label{tb:fine-tune_merge_abl}
\vskip 0.15in
\begin{center}
\begin{small}
\begin{sc}
\begin{tabular}{cc|ccc}
\toprule
model    & pos   & full            & random          & popular         \\ 
\midrule
baseline &       & 0.135           & 0.5573          & 0.0426          \\
(cl4srec)  &       & (0.0585)          & (0.0513)          & (\underline{0.0466}) \\
duorec   & sup   & 0.1346          & 0.5547          & \underline{0.0454}    \\
         & unsup & 0.1358    & 0.5594    & 0.0445          \\
         & -     & 0.1351          & 0.554           & 0.0423          \\ 
\midrule
\rowcolor[HTML]{EFEFEF} 
fisher (with)   &       & \textbf{0.1386} & \textbf{0.5618} & 0.0428   \\
\rowcolor[HTML]{EFEFEF} 
fisher (w.o.)   &       & \underline{0.1373} & \underline{0.5603} & \textbf{0.0487}   \\
\bottomrule
\end{tabular}
\end{sc}
\end{small}
\end{center}
\vskip -0.1in
\end{table}

\subsection{Visualization of Sorted Probability}\label{appendix:sorted_prob}
 The figure displays the sorted probabilities of the top 50 items for 10 sequences, where single line represents single sequence. The cumulative probability values for sample sizes of 10, 30, and 50 are 0.381, 0.569, and 0.658, respectively. With the exception of a few largest ones, the majority of probabilities approximate $0$.
\begin{figure}[h]
\vskip 0.2in
\begin{center}
\centerline{\includegraphics[width=0.8\columnwidth]{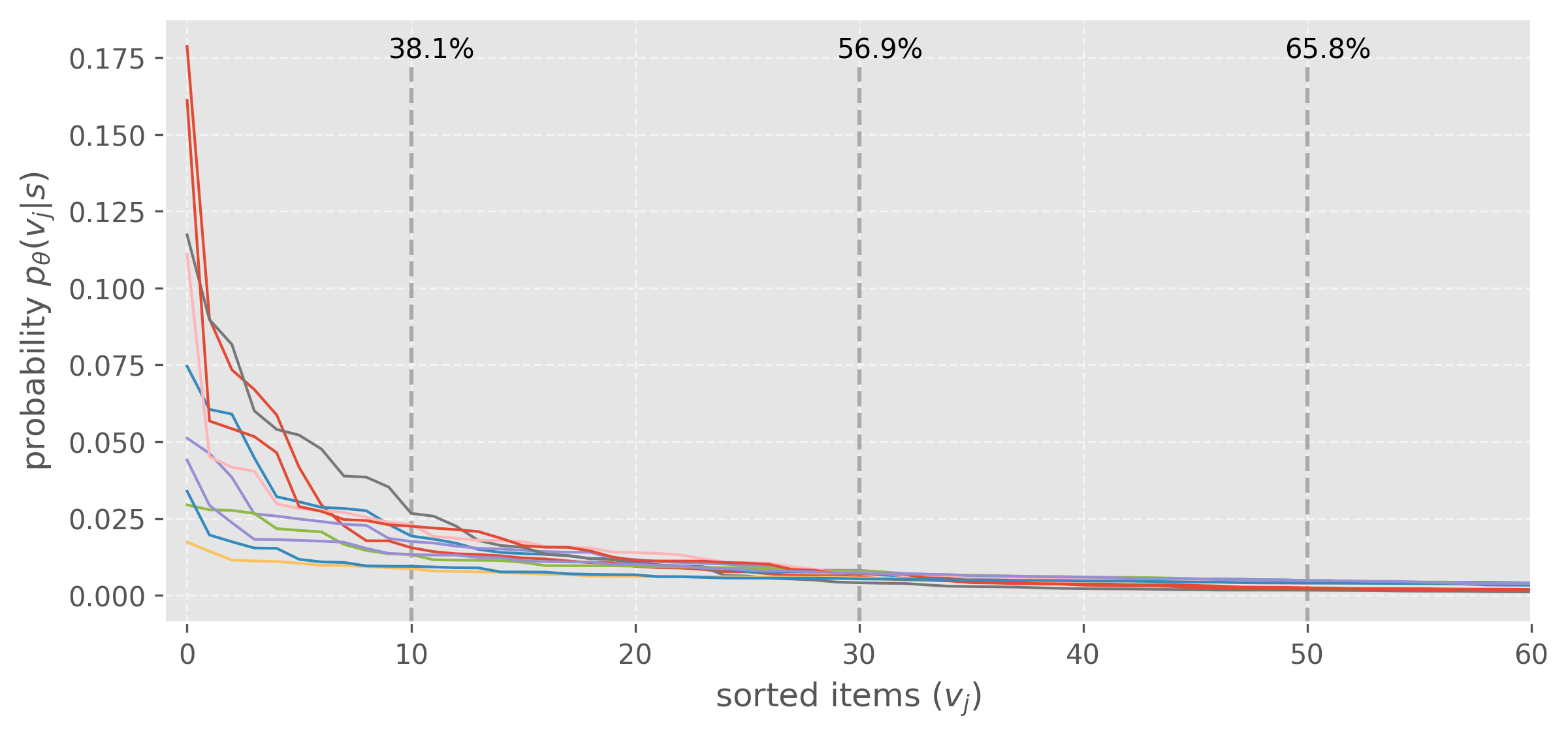}}
\caption{Sorted Probability $p_\theta\left(v_j | s_i\right)$.}
\label{fig:50item_prob}
\end{center}
\vskip -0.2in
\end{figure}


\end{document}